\documentclass[12pt]{article}

\usepackage{xfrac}
\usepackage{natbib}

\usepackage{lipsum}

\newcounter{lastnote}

\title{\bf {\huge{HE0359-3959: an extremely radiating quasar}}}

\author
{M. L. Mart\'{i}nez-Aldama\,$^{1,*}$,  A. Del Olmo\,$^{1}$, P. Marziani\,$^{2}$, C. A. Negrete\,$^{3}$,  \\D. Dultzin\,$^{4}$, M. A. Mart\'{i}nez-Carballo\,$^{1}$\\
\\
\normalsize{$^{1}$Instituto de Astrof\'{i}sica de Andaluc\'{i}a, IAA-CSIC, Granada, Spain}\\
\normalsize{$^{2}$INAF, Osservatorio Astronomico di Padova, Italy}\\
\normalsize{$^{3}$ CONACYT Research Fellow, Instituto de Astronom\'{i}a, UNAM, Mexico}\\
\normalsize{$^{4}$Instituto de Astronom\'{i}a, UNAM, Mexico}\\
\normalsize{$^\ast$Correspondence: maryloli@iaa.es}
}

\date{}

\begin{document} 


\baselineskip24pt


\maketitle

\def\civ{{\sc{Civ}}$\lambda$1549\/}
\def\civonly{{\sc Civ\/}}
\def\hb{{\sc{H}}$\beta$}
\def\hbbc{{\sc{H}}$\beta_{\rm BC}$\/}
\def\caii{{Ca {\sc ii}}}
\def\feiiq{\rm Fe{\sc ii}$\lambda$4570\/\AA\ }
\def\siiii{Si{\sc iii]}$\lambda$1892\/}
\def\heii{He{\sc ii}$\lambda$1640\/}
\def\aliii{Al{\sc iii}$\lambda$1860\/}
\def\siiv{Si{\sc iv}$\lambda$1397\/}
\def\ciii{{\sc{Ciii]}}$\lambda$1909\/}
\def\oiii{{\sc{[Oiii]}}$\lambda\lambda$4959,5007\/}
\def\feii{{Fe\sc{ii}}\/}
\def\rfe{$\mathrm{R_{\rm FeII}}$}
\def\LLedd{L/L$\mathrm{_{Edd}}$}
\def\kms{km s$^{-1}$}
\def\ergs{erg s$^{-1}$}
\def\niv{{N {\sc iv}}$\lambda$1240\/}
\def\nh{n$\mathrm{_H}$}
\def\rblr{r$\mathrm{_{BLR}}$}
\def\mbh{M$\mathrm{_{BH}}$}
\def\mo{M$\mathrm{_\odot}$}
\def\feii{Fe{\sc ii}}
\def\feiii{Fe{\sc iii}}
\def\ufl{erg s$^{-1}$ cm$^{-2}$ \AA$^{-1}$}


\begin{abstract}
  { We present a multiwavelength spectral study of the quasar HE0359-3959, which has been identified as an extreme radiating source at intermediate redshift ($z$=1.5209). Along the spectral range, the different ionic species give information about the substructures in the broad line region. {The presence of a powerful outflow with an extreme blueshifted velocity of $\sim$--6000{$\pm$500} \kms\ is shown in the \civ\ emission line.} A prominent blueshifted component is also associated with the 1900\AA\ blend, resembling the one observed in \civ. We detect a strong contribution of very the low--ionization lines, \feii\ and Near-Infrared \caii\ triplet.{We find that the physical conditions for the low, intermediate and high--ionization emission lines are different, which indicate that the emission lines are emitted in different zones of the broad line region. The asymmetries shown by the profiles reveal different forces over emitter zones. The high--ionization region is strongly dominated by radiation forces, which also affect the low and intermediate--ionization emitter region, commonly governed by virial motions.} These results support the idea that highly radiating sources host a slim disk.} 
\end{abstract}



\section{Extreme Population A sources along the 4DE1 Main Sequence}

The 4D Eigenvector 1 (4DE1) parameter space offers a formalism to distinguish and classify type 1 Active Galactic Nuclei (AGN) considering their spectral properties \citep{SUL00a, SUL00b}. The {Full Width at Half Maximum} (FWHM) of \hb\ broad component (H$\beta_{\rm {BC}}$), the strength of optical \feii\ blend at 4570\AA\ described by the ratio \rfe = I(\feii)/I(\hbbc), the velocity shift of the \civ\ profile, and soft X-ray photon index ($\Gamma \mathrm{_{soft}}$), provide four observationally independent dimensions of the Eigenvector 1. In the 4DE1 optical plane, the type 1 AGN occupy a well defined sequence, driven mainly by the Eddington ratio, \LLedd. Along this sequence we observe a variation of the physical parameters and orientation. Then, 4DE1 could be revealing an evolution sequence for type 1 AGN \citep{SUL00a, MAR10, ZAM10}. For more information about the 4DE1 and {update of} results, see Marziani et al. 2017 in this volume. \\

{Using the 4DE1 we identify two populations} with different spectral features: A and B. Population A has a FWHM(\hb$_{\rm {BC}}$)$\leq$4000 \kms. It shows large blue asymmetries in the high-ionization lines like \civ, and it is majority populated by radio quiet sources. In contrast, population B shows a FWHM(\hb$\rm {_{BC}}$)$>$4000 \kms\ and it is mostly composed of radio-loud sources \citep{SUL02, ZAM10}. Each population can be divided into small bins with $\Delta$FWHM(\hb$\rm {_{BC}}$)=4000 \kms\ and $\Delta$\rfe=0.5, defining  subpopulations shown in the Figure \ref{fig:4DE1}. In this paper we focus in the subpopulation A3 and A4 (\rfe$>$1), which have been identified as highly radiating sources  \citep[xA,][]{MAR14}. These kind of sources show high Eddington ratios (\LLedd$>$0.2) probably produced by a slim disk, which is geometrically and optically thick and it could be formed in an advection-dominated accretion flow \citep{ABRA88, ABRA14}. \\ 

We have found selection criteria to identify the xA sources based on the 4DE1 formalism. In the optical region they show a \rfe$>$1 (high intensity of \feii) and in the UV range \aliii/\siiii$\geq$0.5 and \ciii/\siiii$\leq$1.0 \citep{MAR14}. Also, they show strong blueshifted components associated with the high ionization lines, for example in \civ\ emission line, indicating the presence of outflows. More details about the xA sources behavior can be found in Mart{\'{\i}}nez-Aldama et al. of this volume.

\subsection{HE0359-3959: an extreme xA source}

\begin{figure}[htp!]
\begin{center}
\includegraphics[width=8cm]{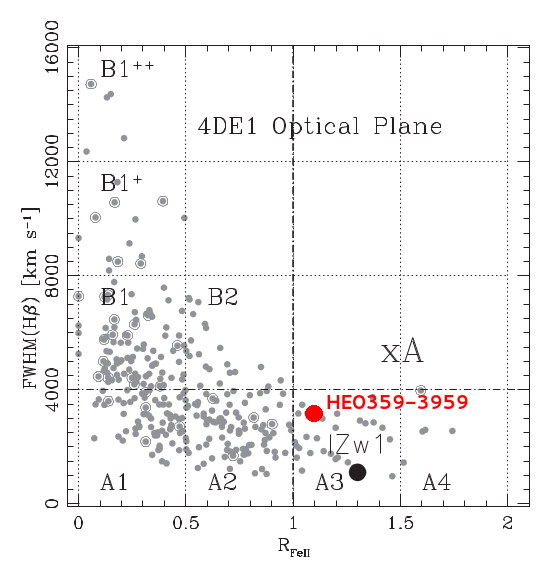} 
\end{center}
\caption{4DE1 Optical Plane reproduced from \citet{MAR14}. Grey points correspond to the sample of 470 bright low-$z$ QSOs from \citet{ZAM10}. The plane is divided in bins according to \citet{SUL02}. Extreme accretor population A sources (xA) are located in A3 and A4 bins. The black dot indicates the position of 1 Zw 1, the prototype of low-$z$ xA sources. And, the red dot marks the location of HE0359-3959, {an extreme} xA source with high-$z$.}
\label{fig:4DE1}
\end{figure}

In our extreme luminosity Hamburg-ESO sample \citep{MAR09, SUL17}, we have identified four cases of highly radiating quasars that show an extreme behavior, i.e., a high Eddington ratio and a strong blue asymmetry (c($\sfrac{1}{2}$)$<$--4000 \kms; centroid at half intensity) in the \civ\ profile \citep{SUL17}. The most extreme case corresponds to the quasars HE0359-3959, with $z$=1.5209, log(L$\mathrm{_{bol}}$)=47.6 \ergs\ and a \rfe=1.12. It is cataloged as an A3 source (see Figure \ref{fig:4DE1}). \\

{The aim of this paper is to analyze the spectral behavior of an extreme xA source, the quasar HE0359--3959. We performed multicomponent fits in a wide spectral range: UV, optical and Near--Infrared (Section \ref{sec:obsdat}); which gives us information about the dynamics and the physical conditions of the broad line region (BLR) (Section \ref{sec: results}). In Section \ref{sec:conclusions}, we summarize the main results of our work.}

\section{Observations, data reduction and multicomponent fitting}
\label{sec:obsdat}

\subsection{Observations and data reduction}
\label{sec:obs}

Ultraviolet (UV), optical and Near--Infrared spectra were observed with the Very Large Telescope (VLT-ESO). Optical and Near--Infrared spectra were obtained with the Infrared Spectrometer And Array Camera (ISAAC; decommissioned in 2013) {using a slit of 0.6$''$. The near--infrared spectrum was observed in 2010 in the K band with a total exposure time of 1120 seconds. The optical spectrum was observed in 2004 in the J band with a total exposure time of 3600 seconds}.  {For the ultraviolet spectrum we used the Focal Reducer and low dispersion Spectrograph (FORS1) {\and a slit of 1.0$''$ with a total exposure time of 1440 seconds. It was observed in 2008.} The data reduction was done using the {\sc iraf} package. The procedures followed are explained in \citet{MAR09}, \citet{MAR15} and \citet{SUL17}. }

\subsection{Multicomponent fits}
\label{sec: multifits}

\begin{figure}[htp!]
\begin{center}
\includegraphics[width=18cm]{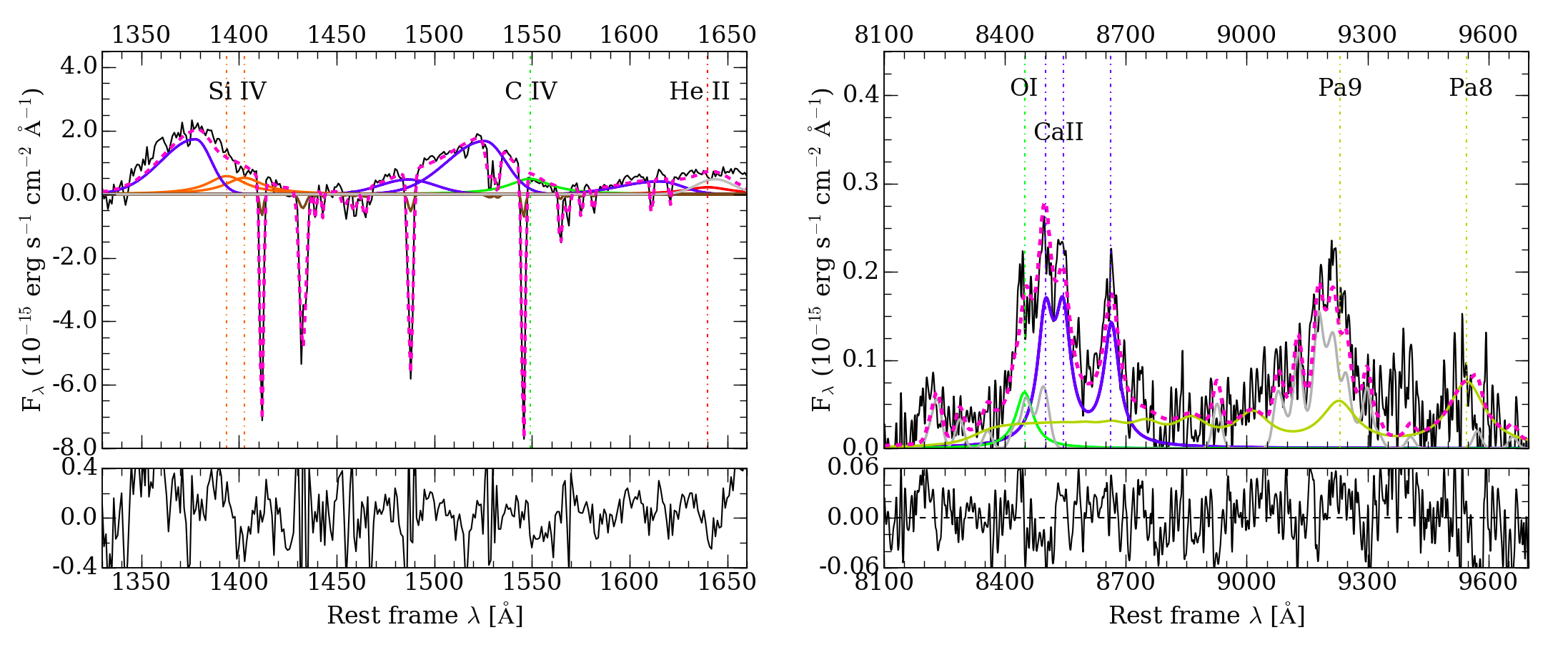}
\end{center}
\caption{ {Multicomponent analysis on the UV, optical and NIR spectra of HE0359-3959 after continuum subtraction. {\sc Top panels}: In the left side is shown the UV spectrum, while in the right one is shown the near--infrared spectrum. The different components (broad (BC), BLUE, and narrow (NC)) in the line fitting are specified in each panel. Vertical lines indicate the rest-frame obtained through H$\beta_{{\mathrm {NC}}}$. The gray line marks the \feii\ contribution. The vertical scale represents the relative flux in units of 10$^{-15}$ \ufl. {\sc Bottom panel}: Residuals of the fittings. The horizontal scale is the radial velocity shift in \kms. In all the panels the horizontal scale represents the rest--frame wavelength in \AA.}}
\label{fig:spectra}
\end{figure}

{We perform multicomponent fits using {\sc specfit}, an {\sc iraf} routine \citep{KRI94} to get the information of the most important emission lines. In each spectral range we fit a local continuum. The FWHM of all the broad components (BC) for \hb, \aliii, \siiii, \civ\ and \siiv\ was taken equal. In the Figure \ref{fig:spectra}, we present the multicomponents fits after continuum subtraction, for the \civ\ and \caii\ triplet range. The rest of the fits will be shown in an upcoming paper. }

\section{Results}
\label{sec: results}

\subsection{Multiwavelength analysis}
\label{sec: analysis}

{Low--ionization lines (LIL) have an ionization potential (IP) $\leq$20 eV. The H$\beta$ line is the prototype of LIL. In population A3 and A4 sources \hb\ has associated a blueshifted component \citep{BACH04}. In the case of HE0359-3959, the blueshifted component has a contribution to the total flux of $\sim$9$\%$, and shows a centroid a half intensity of  c($\sfrac{1}{2}$)$\approx$--500{$\pm$70} \kms.}\\

{The \feii\ (IP$\sim$16 eV) has an important contribution in the optical and near--infrared regions. To reproduce it we used the templates modeled by \citet{MAR09} and \citet{GAR12} for the optical and near--infrared ranges, respectively. Several works have found \citep{JOL89, PERS88, FERPER89, DUL99, MAR15} a close relationship between the \feii\ and the NIR \caii\ $\lambda$8498,$\lambda$ 8542 and $\lambda$8662 \AA\ triplet. This relation is very well appreciable in this object: as well as the optical \feii\ is strong, the NIR \caii\ triplet also is. It is the first time where we observe the \caii\ triplet lines isolated at high redshift.} Strong intensities of both ions imply an extremely low-ionization degree ($U<$10$^{-2}$; $U$: {ionization parameter}) and a high density (\nh$\sim$10$^{11-13}$ cm$^{-3}$) \citep{BAL04, MAT07, MAR15}.\\

{In the UV region, the 1900\AA\ blend is formed by two intermediate--ionization lines (IIL; IP$\sim$20--40 eV), \aliii\ and \siiii, which are accompanied by  \ciii\ and some \feiii\ transitions. In this blend we appreciate a blueshifted component. This component should be most likely associated with \aliii. Respect to \aliii, the blueshifted component has a contribution of the total profile of 60$\%$. The centroid a half intensity is c($\sfrac{1}{2}$)$\approx$--3200{$\pm$250} \kms, which indicates the presence of an outflow generated by radiation forces presented in the intermediate--ionization lines \citep{MAR17}.} On the other hand, considering the high intensity of \aliii, \siiii, \caii\ and \feii, it could suggest a possible chemical enrichment of the BLR \citep{JUA09}. \\

High ionization lines (HIL; IP$>$40 eV), \civ, \heii\ and \siiv, show a prominent blueshifted component. We find that the blue component has a contribution of 76$\%$, 62$\%$ and 57$\%$ to the total flux of \civ, \heii\ and \siiv\ respectively. The \civ\ reaches c($\sfrac{1}{2})\sim$--6000{$\pm$500} \kms, while \heii\ and \siiv\ c($\sfrac{1}{2})\sim$--4000{$\pm$550}  \kms. {The velocities reached} are ones of the highest found in the literature \citep{RICH11, COAT16, SUL17}. Then, it indicates that the full profile is dominated by an outflow and suggests the disk plus wind scenario \citep{GASK82, RICH02, RICH11}.

\subsection{Physical properties of HE0359-3959}
\label{sec:physprop}

In order to study the physical properties of the quasar HE0359-3959, we built a grid of photoionization simulations using the {\sc CLOUDY} code \citep{FER98, FER13}. For our simulations we considerer a Mattews and Ferland continuum \citep{MATFER87}, a plane-parallel geometry, a metallicity 5Z$\odot$ with an overabundance of Al and Si with respect to carbon (by a factor of three), and a column density of N${\rm _c}$=10$^{23}$ cm$^{-2}$. {See \citet{NEG12} for more details.} Our simulations span the density range 7.00$\leq$log(\nh)$\leq$14.00 and --4.5$\leq$log($U$)$\leq$0.00 for the ionization parameter, in intervals of 0.25 dex. More details about the {\sc cloudy} simulations can be found in \citet{NEG14}. Using the UV lines, we define three groups of diagnostic ratios::

\begin{itemize}

\item The flux ratio \aliii/\siiii\ is a useful density diagnostic.

\item The flux ratio \siiv/\siiii\ for the ionization parameter.

\item The flux ratio \civ/\siiv\ is mainly sensitive to the relatives abundances of C and Si. \\

\end{itemize}

\begin{figure}[htp!]
\begin{center}
\includegraphics[width=10cm]{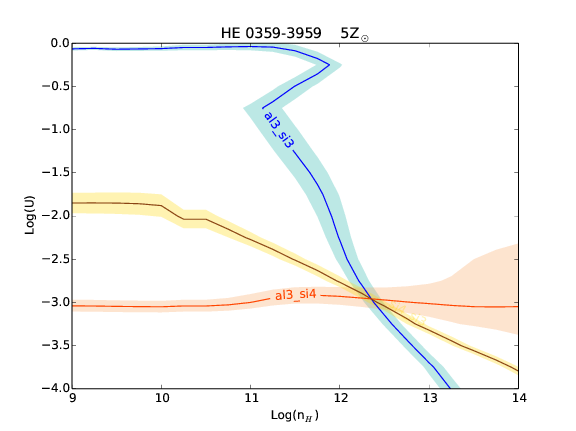} 
\end{center}
\caption{Isocontours for HE0359-3959 with 5Z$_\odot$ and an overabundance of Al and Si. The blue line indicates the flux ratio \aliii/\siiii, the yellow one indicates the ratio \siiv/\siiii\ and the orange corresponds to the \aliii/\siiv. Shadows associated with each line indicate the error. The flux ratios are intersected in \nh$\cdotp U$=9.27$\pm$0.39}.
\label{fig:cloudy}
\end{figure}

In Figure \ref{fig:cloudy} is shown the result of the simulations. We obtained that the flux ratios are intersected in log(\nh)=12.32 cm$\mathrm{^{-3}}$ and log($U$)=--2.95. Compared to not highly radiating AGNs \citep{NEG13}, this source shows a high density and a low ionization parameter, which marks a different behavior in the BLR, probably causing by the slim disk hosted in these kind of sources. Taking into account the high intensity of \aliii, \feii\ and \caii\ we conclude that effectively the low--ionization emitter zone has a high density and low--ionization parameter. \\

\citet{NEG12} proposed a new method to determine the size of the BLR (\rblr) and the black hole mass (\mbh) based on the product \nh$\cdotp U$ and independently of redshift. This method gives similar results to the obtained from the classical methods such as reverberation mapping at low--$z$ \citep{NEG14}. Knowing the product of \nh$\cdotp U$ obtained from the {\sc cloudy} simulations, we  compute the size of the BLR (\rblr) and considering the FWHM of the broad components as the velocity dispersion, we can get the black hole mass (\mbh) and the Eddington ratio. The size of the BLR is log(\rblr)=18.37$\pm$0.04 cm and the black hole mass is log(\mbh)=9.52$\pm$0.41 \mo. These values are in agreement with the ones found for a large xA sample at high-redshift (Mart{\'{\i}}nez-Aldama et al. in prep.). \\

The Eddington ratio for this source is \LLedd=0.74$\pm$0.11. Considering that it shows a c($\sfrac{1}{2}$)$\sim$--6000$\pm$500 \kms\ for \civ, we confirm the directly proportional relation between c($\sfrac{1}{2}$) and \LLedd. Indicating that \LLedd could be the driver of the outflows \citep{SUL17}.

\section{Conclusions}
\label{sec:conclusions}

{The information given by the multiwavelength analysis indicates that in HE0359--3959 there is coexistence of substructures in the broad line region. Low and intermediate--ionization regions, where \hb, \aliii\ and \siiii\ are emitted, are dense (\nh$\sim$10$^{11-12}$ cm$^{-3}$) and optically thick ($U\sim$10$^{-2.5}$). They are mainly governed by virial motions and the presence of a blueshifted component indicates the influence of radiation forces. On the other hand, according to \citet{MAR10} the high--ionization region is less dense (\nh$\sim$10$^{10}$ cm$^{-3}$, $U\sim$10$^{-1}$), pointing out a difference with the physical conditions shown by the low and intermediate--ionization lines.\\

High ionization lines are dominated by strong radiation forces, producing outflows in high--ionization lines like \civ, \heii\ and \siiv. The high Eddington ratio value suggests the presence of a slim optically thick disk which could be related to the extreme outflow properties observed in HE0359-3959. The presence of strong outflows has been related with the co--evolution of the active galactic nuclei and the host galaxy. } \\ \\

{\small {\textit {Acknowledgements.} M.L.M.A acknowledge the postdoctoral grant from CONACyT. M.L.M.A., A.d.O. and M.A.M.C. acknowledge financial support from Spanish Ministry for Economy and Competitiveness through grants AYA2013-42227-P and AYA2016-76682-C3-3-1-P.}}




\begin{thebibliography}{}

\bibitem[Abramowicz et al.(1988)]{ABRA88} 
 {\sc Abramowicz}, M.~A., {Czerny}, B., {Lasota}, J.~P. and {Szuszkiewicz}, E., \ 1988,  Astrophys. J., 332, 646-658  

\bibitem[Abramowicz $\&$ Straub (2014)]{ABRA14} 
 {\sc Abramowicz}, M.~A. $\&$ {Straub}, O., \ 2014, 9 
 
\bibitem[Bachev et al.(2004)]{BACH04} 
 {\sc Bachev}, R., {Marziani}, P., {Sulentic}, J.~W., {Zamanov}, R., {Calvani}, M. and {Dultzin-Hacyan}, D.,\ 2004,  Astrophys. J., 617, 171-183 
   
\bibitem[Baldwin et al.(2004)]{BAL04} 
 {\sc Baldwin}, J.~A., {Ferland}, G.~J., {Korista}, K.~T., {Hamann}, F. and {LaCluyz{\'e}}, A., \ 2004, Astrophys. J., 615, 610-624

\bibitem[Coatman et al.(20016)]{COAT16} 
 {\sc Coatman}, L. and {Hewett}, P.~C. and {Banerji}, M. and {Richards}, G.~T., \ 2016, Mon. Not. Roy. Astron. Soc., 461, 647-665
   
\bibitem[Dultzin et al.(1999)]{DUL99} 
 {\sc Dultzin-Hacyan}, D., {Taniguchi}, Y. and {Uranga}, L., \ 1999, PASP, 175, 303 

\bibitem[Ferland $\&$ Persson(1989)]{FERPER89} 
 {\sc Ferland}, G.~J. $\&$ {Persson}, S.~E., \ 1989,  Astrophys. J., 347, 656-673
 
\bibitem[Ferland et al.(1998)]{FER98} 
 {\sc Ferland}, G.~J., {Korista}, K.~T., {Verner}, D.~A., {Ferguson}, J.~W., {Kingdon}, J.~B. and {Verner}, E.~M., \ 1998, PASP, 110, 761-778
 
\bibitem[Ferland et al.(2009)]{FER09} 
 {\sc Ferland}, G.~J., {Hu}, C., {Wang}, J.-M., {Baldwin}, J.~A., {Porter}, R.~L., {van Hoof}, P.~A.~M. and {Williams}, R.~J.~R., \ 2009, Astrophys. J., 707, L82-L86
  
\bibitem[Ferland et al.(2013)]{FER13} 
 {\sc Ferland}, G.~J., {Porter}, R.~L., {van Hoof}, P.~A.~M., {Williams}, R.~J.~R., {Abel}, N.~P., {Lykins}, M.~L., {Shaw}, G., {Henney}, W.~J. and {Stancil}, P.~C., \ 2013, RMxAA, 49, 137-163

\bibitem[Garc{\'{\i}}a-Rissmann et al.(2012)]{GAR12} 
 {\sc Garc{\'{\i}}a-Rissmann}, A., {Rodr{\'{\i}}guez-Ardila}, A. and {Sigut}, T.~A.~A. and {Pradhan}, A.~K., \ 2012, Astrophys. J., 751, 7
  
\bibitem[Gaskell(1982)]{GASK82} 
 {\sc Gaskell}, C.~M., \ 1982,  Astrophys. J., 263, 79-86

\bibitem[Joly(1989)]{JOL89} 
 {\sc Joly}, M., \ 1989, Astron. Astrophys., 208, 47-51

\bibitem[Juarez et al.(2009)]{JUA09} 
 {\sc Juarez}, Y., {Maiolino}, R., {Mujica}, R., {Pedani}, M., {Marinoni}, S., {Nagao}, T., {Marconi}, A. and {Oliva}, E., \ 2009,  Astron. Astrophys., 494, L4-L28
  
\bibitem[Kriss(1994)]{KRI94} 
{\sc Kriss}, G., \ 1994, Astronomical Data Analysis Software and Systems, 3, 437 

\bibitem[Mart{\'{\i}}nez-Aldama et al.(2015)]{MAR15} 
 {\sc Mart{\'{\i}}nez-Aldama}, M.~L., {Dultzin}, D., {Marziani}, P., {Sulentic}, J.~W., {Bressan}, A., {Chen}, Y. and {Stirpe}, G.~M., \ 2015, Astrophys. J. S., 217, 3

\bibitem[Marziani et al.(2001)]{MAR01} 
 {\sc Marziani}, P., {Sulentic}, J.~W., {Zwitter}, T., {Dultzin-Hacyan}, D. and {Calvani}, M., \ 2001,  Astrophys. J., 558, 553-560
  
\bibitem[Marziani et al.(2009)]{MAR09} 
 {\sc Marziani}, P., {Sulentic}, J.~W., {Stirpe}, G.~M., {Zamfir}, S. and {Calvani}, M., \ 2009, Astron. Astrophys., 495, 83-112 

\bibitem[Marziani et al.(2010)]{MAR10} 
 {\sc Marziani}, P., {Sulentic}, J.~W., {Negrete}, C.~A., {Dultzin}, D., {Zamfir}, S. and {Bachev}, R., \ 2010, 409, 1033-1048
 
\bibitem[Marziani $\&$ Sulentic(2014)]{MAR14} 
 {\sc Marziani}, P. $\&$ {Sulentic}, J.~W., \ 2014,  Mon. Not. Roy. Astron. Soc., 442, 1211-1229  
 
\bibitem[Marziani et al.(2017)]{MAR17} 
 {\sc Marziani}, P., {Del Olmo}, A.,  {Martínez-Aldama}, M. ~L.,  {Dultzin}, D., {Negrete}, C., {Bon}, E., {Bon}, N. and {D'Onofrio}, M., \ 2017, Atoms, 5, 33-47
  
\bibitem[Matsuoka et al.(2007)]{MAT07} 
 {\sc Matsuoka}, Y., {Oyabu}, S., {Tsuzuki}, Y. and {Kawara}, K., \ 2007, Astrophys. J., 663, 781-798

\bibitem[Matsuoka et al.(2008)]{MAT08} 
 {\sc Matsuoka}, Y., {Kawara}, K. and {Oyabu}, S., \ 2008, Astrophys. J., 673, 62-68
 
\bibitem[Mathews $\&$ Ferland(1987)]{MATFER87} 
 {\sc Mathews}, W.~G. $\&$ {Ferland}, G.~J., \ 1987,  Astrophys. J., 323, 456-467

\bibitem[Negrete et al.(2012)]{NEG12} 
 {\sc Negrete}, C.~A., {Dultzin}, D., {Marziani}, P. and {Sulentic}, J.~W., \ 2012,  Astrophys. J., 757, 62 
 
\bibitem[Negrete et al.(2013)]{NEG13} 
 {\sc Negrete}, C.~A., {Dultzin}, D., {Marziani}, P. and {Sulentic}, J.~W.,  \ 2013, Astrophys. J. 771, 31
 
\bibitem[Negrete et al.(2014)]{NEG14} 
 {\sc Negrete}, C.~A., {Dultzin}, D., {Marziani}, P. and {Sulentic}, J.~W., \ 2014, Astrophys. J., 794, 95

\bibitem[Persson(1988)]{PERS88} 
 {\sc Persson}, S.~E., \ 1988,  Astrophys. J., 330, 751-765

\bibitem[Richards et al.(2002)]{RICH02} 
 {\sc Richards}, G.~T., {Vanden Berk}, D.~E., {Reichard}, T.~A., {Hall}, P.~B., {Schneider}, D.~P., {SubbaRao}, M.,	{Thakar}, A.~R. and {York}, D.~G., \ 2002, Astrophys. J., 124, 1-17

\bibitem[Richards et al.(2011)]{RICH11} 
 {\sc Richards}, G.~T., {Kruczek}, N.~E., {Gallagher}, S.~C., {Hall}, P.~B., {Hewett}, P.~C., {Leighly}, K.~M., {Deo}, R.~P., {Kratzer}, R.~M. and {Shen}, Y., \ 2011, Astrophys. J., 141 , 167
  
\bibitem[Shen $\&$ Ho(2014)]{SHEN14} 
 {\sc Shen}, Y. $\&$ {Ho}, L.~C., \ 2014, Nature, 513, 210-213
  
\bibitem[Shin et al.(2013)]{SHI13} 
 {\sc Shin}, J., {Kim}, S.~S. and {Yoon}, S.-J. and {Kim}, J., \ 2013, Astrophys. J., 762 , 135

\bibitem[Sulentic et al.(2000a)]{SUL00a} 
{\sc Sulentic}, J.~W., {Marziani}, P. and {Dultzin-Hacyan}, D., \ 2000a, Annu. Rev. Astron. Astrophys., 38, 521-571 

\bibitem[Sulentic et al.(2000b)]{SUL00b} 
{\sc Sulentic}, J.~W., {Zwitter}, T., {Marziani}, P. and {Dultzin-Hacyan}, D., \ 2000b, Astrophys. J., 536, L5-L9 

\bibitem[Sulentic et al.(2002)]{SUL02} 
{\sc Sulentic}, J.~W., {Marziani}, P., {Zamanov}, R., {Bachev}, R., {Calvani}, M. and {Dultzin-Hacyan}, D., \ 2002, Astrophys. J., 566, L71-L75 

 \bibitem[Sulentic et al.(2007)]{SUL07} 
{\sc Sulentic}, J.~W., {Bachev}, R., {Marziani}, P., {Negrete}, C.~A. and {Dultzin}, D., \ 2007,  Astrophys. J., 666, 757-777

\bibitem[Sulentic et al.(2011)]{SUL11} 
{\sc Sulentic}, J., {Marziani}, P. and {Zamfir}, S., \ 2011, Baltic Astronomy, 20, 427-434 

\bibitem[Sulentic et al.(2017)]{SUL17}  
 {\sc Sulentic}, J.~W., {Del Olmo}, A., {Marziani}, P., {Mart{\'{\i}}nez-Carballo}, M. A., {D'Onofrio}, M., {Dultzin}, D., {Mart{\'{\i}}nez-Aldama}, M. ~L., {Negrete}, C., {Stirpe}, G. ~M. and {Zamfir}, S., \ 2017, Astron. $\&$ Astrophys., 608, 122 

\bibitem[Vietri(2017)]{VIE17} 
{\sc Vietri}, G., \ 2017,  American Astronomical Society Meeting, 229, 302
  
\bibitem[Zamfir et al.(2010)]{ZAM10} 
{\sc Zamfir}, S., {Sulentic}, J.~W., {Marziani}, P. and {Dultzin}, D., \ 2010, Mon. Not. Roy. Astron. Soc., 403, 1759-1786  

\end{thebibliography}
\end{document}